\documentclass[aps,showpacs,preprintnumbers,amsmath,amssymb,superscriptaddress,twocolumn]{revtex4}

\usepackage{graphicx,color}
\usepackage{dcolumn}
\usepackage{bm}
\usepackage[active]{srcltx}

\begin{document}

\title{Chiral criticality in doped Mn$_{1-y}$Fe$_y$Si compounds.}

\author{S. V. Grigoriev}
\author{E. V. Moskvin}
\author{V. A. Dyadkin}
\affiliation{Petersburg Nuclear Physics Institute, Gatchina, 188300 Saint-Petersburg, Russia}

\author{D. Lamago}
\affiliation{Laboratoire L\'eon Brillouin, CEA Saclay, F-91191 Gif-sur-Yvette Cedex, France}
\affiliation{Karlsruher Institut f\"ur Technologie, Institut f\"ur Festk\"orperphysik, P.O. box 3640, D-76021 Karlsruhe, Germany}

\author{Th. Wolf}
\affiliation{Karlsruher Institut f\"ur Technologie, Institut f\"ur Festk\"orperphysik, P.O. box 3640, D-76021 Karlsruhe, Germany}

\author{H. Eckerlebe}
\affiliation{Helmholtz Zentrum Geesthacht, 21502 Geesthacht, Germany}

\author{S. V. Maleyev}
\affiliation{Petersburg Nuclear Physics Institute, Gatchina, 188300 Saint-Petersburg, Russia}

\date{\today}

\begin{abstract}
The critical spin fluctuations in doped compounds Mn$_{1-y}$Fe$_y$Si have been studied by means of  ac-susceptibility measurements, polarized neutron small angle scattering and spin echo spectroscopy. It is shown that these compounds undergo the transition from the paramagnetic to helimagnetic phase through continuous, yet well distinguishable crossovers: (i) from paramagnetic to partially chiral, (ii) from partially chiral to highly chiral fluctuating state. The crossover points are identified on the basis of combined analysis of the temperature dependence of ac-susceptibility and polarized SANS data. The whole transition is marked by two inflection point of the temperature dependence of ac-susceptibility: the upper one corresponds to the crossover to partially chiral state at $T^*$, where the inverse correlation length $\kappa \approx 2 k$, the lower one corresponds to the transition to the spin helix structure.  The intermediate crossover to the highly chiral phase is observed at the inflection point $T_k$ of the first derivative of ac-susceptibility, where $\kappa \approx k$. The temperature crossovers to the highly chiral fluctuating state is associated with the enhancing influence of the Dzyaloshinskii-Moria interaction close to $T_c$. 
\end{abstract}



\maketitle
{\it Introduction.} The cubic B20-type (space group P2$_1$3) doped compounds Mn$_{1-y}$Fe$_y$Si with $y \in [0;0.15]$ order in a spin helix structure with a small propagation vector $0.36 \leq k \leq 0.70$~nm$^{-1}$ \cite{NishiharaPRB84_v30, Grigoriev09PRB_v79}. The spin helix structure is well interpreted within the Bak-Jensen (B-J) hierarchical model \cite{BakJPCM80_v13}. 
The hierarchy implies that the spin helix appears as a result of the competition between the ferromagnetic spin exchange and antisymmetric Dzyaloshinskii-Moriya interaction (DMI) caused by the lack of inverse symmetry in arrangement of magnetic Mn atoms \cite{BakJPCM80_v13,Dzyaloshinskii64ZETF,NakanishiSSC80_v35}. The weak anisotropic exchange interaction fixes the direction of the spin helix propagation. 

The nature of the magnetic phase transition even in pure MnSi remains a subject of intensive debates \cite{RoessliPRL02_v88, GrigorievPRB05_v72, RoesslerNature06_v442, StishovPRB07_v76, PappasPRL09_v102, GrigorievPRB10_v81_2}. The detailed study of the critical neutron scattering in pure MnSi had been reported in \cite{GrigorievPRB05_v72}. It was shown that the scattering intensity of polarized neutrons above $T_c=28.8$~K looks like half-spheres centred at the origin and oriented along the incident neutron polarization. This scattering was referred to the purely chiral spiral spin fluctuations with the wavevector ${\bf k}$ of the random orientation. The high degree of chirality of the fluctuations emphasises  a relative strength of the DMI. The sum of the intensities for two opposite polarizations forms an anisotropic sphere close to $T_c$ with weak spots, which below $T_c$ transform into the Bragg peaks assigned to the helical structure. The mean-field theory of the critical fluctuations based on the (B-J) model was proposed and shown to describe the experimental data very well.

Later the closer look had revealed existence of a 100\% chiral spin fluctuations within one kelvin above $T_{c}$ \cite{PappasPRL09_v102}. It was suggested that this region may be a candidate for the spontaneous fluctuating skyrmion phase \cite{PappasPRL09_v102} as was assumed in \cite{RoesslerNature06_v442}. Later in \cite{GrigorievPRB10_v81_2} we demonstrated in more details, compare to \cite{GrigorievPRB05_v72}, that existing experimental data are in good qualitative agreement with the conventional mean-field theory, which catches principal symmetry features of the problem. To emphasize the generality of our approach  we note that the similar expression for the neutron cross section had been obtained for  MnSi under applied pressure (see Eq.10 in Ref. \cite{TewariPRL06_v96}).

Moreover, we stress that the mean-field theory \cite{GrigorievPRB05_v72} accounting the modifications of ferromagnetic fluctuations by DMI does not imply any texture of the fluctuating object (spiral or skyrmion).  This theory describes the correlation length increasing on approaching $T_{c}$. Therefore the expressions derived in \cite{GrigorievPRB05_v72,GrigorievPRB10_v81_2} for the susceptibility, the neutron cross section and for the correlation functions must be valid for the fluctuations of both spiral or skyrmion textures. 

The magnetic properties of the doped compounds Mn$_{1-y}$Fe$_{y}$Si as well as Mn$_{1-y}$Co$_{y}$Si were recently studied by the measurements of the magnetization and ac-susceptibility \cite{BauerPRB10_v82}. Remarkably that high field magnetization measurements demonstrated that the temperature dependence of the saturation magnetization of the doped compounds is well described by $m_s^2 = m_{s,0}^2(1-T^2/T_m^2)$, where $T_m$ (denoted as $T_c$ in \cite{BauerPRB10_v82}) is the critical temperature at which magnetization of the compound is extrapolated to zero. $T_m$ appears to be proportional to $(1-\sqrt{x/x_c})$ with $x_c= 0.192$ for the Mn$_{1-x}$Fe$_{x}$Si compounds. These findings confirm the conclusions made in  \cite{Grigoriev09PRB_v79} that ferromagnetic exchange representing the strongest energy scale of the system leads to the underlining ferromagnetic quantum critical point at $x_c$.

\begin{figure}[hbtp]
\begin{center}\includegraphics[width=8.5cm]{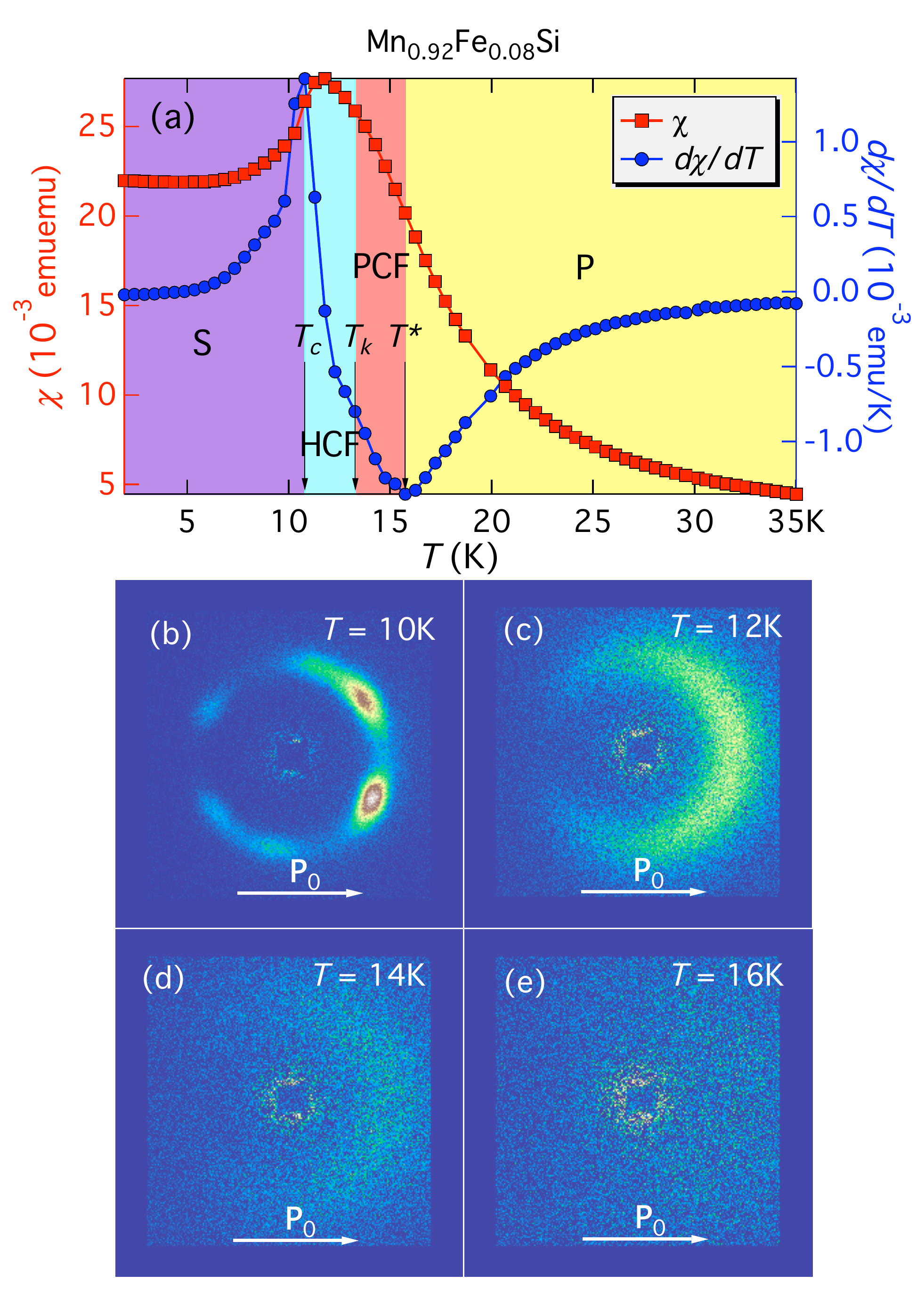}
\caption{The sample with concentration $y=0.08$. (a) --- dependencies $\chi(T)$ and $d\chi/dT(T)$ in the magnetic field $H=50$~mT; four different temperature regions (S, HCF, PCF and P --- see explanations in text) are shown. Four SANS maps corresponding to the shadowed/coloured areas on the temperature scale are: (b) $T<T_{c}$, (c)  $T_{c}\le T\le T_{k}$, (d)  $T_{k}\le T\le T^{*}$, and (e)  $T>T^{*}$.}
\label{fig:PPMS+SANS}
\end{center}
\end{figure}

The DMI modifies critical phenomena notably. At least three specific temperature points can be extracted from the temperature dependence of ac-susceptibility. The typical curve (see Fig.~\ref{fig:PPMS+SANS}(a)) is characterized by the points of inflection of susceptibility at low temperature $T_c$ (transition into the spiral phase) and at high temperature $T^*$ (at the border of paramagnetism). These two points are denoted as $T_{c1}$ and $T_{c2}$ in ref.\cite{BauerPRB10_v82}, respectively. However, the true identification and the nature of these characteristic points is impossible without complimentary data taken by small angle neutron diffraction. In this paper we fill the gap and demonstrate the phenomena occurring in the temperature range between $T_c$ and $T^*$ in the light of neutron scattering.    


{\it AC-susceptibility and polarised neutron scattering measurements.} A series of Mn$_{1-y}$Fe$_y$Si single crystals with $y = 0.0$, 0.08, 0.10, 0.11, 0.12 have been studied using ac-susceptibility measurements at PPMS station in HZB, Berlin, Germany and small angle polarized neutron scattering (SAPNS) at SANS-2 instrument of HZG, Geesthacht, Germany.
The examples of these measurements for the sample Mn$_{0.92}$Fe$_{0.08}$Si are sketched in Fig~\ref{fig:PPMS+SANS}. The temperature dependence of susceptibility, $\chi$, is shown in Fig.~\ref{fig:PPMS+SANS}(a). Here we also plot the first derivative of the susceptibility on the temperature $d\chi/dT$ to emphasize the inflection points $T_c$ and $T^*$ on the $\chi(T)$ dependence. These inflection points divide the temperature scale into the three regions: (i) from low temperatures to maximum of $d\chi/dT$; (ii) between maximum and minimum of $d\chi/dT$; and (iii) from minimum of first derivative $d\chi/dT$ to the higher temperatures (Fig.~\ref{fig:PPMS+SANS}(a)). Yet another characteristic temperature which should be distinquished (denoted as $T_k$) corresponds to the minimum of second derivative on the temperature $d^{2}\chi/dT^{2}$ within the range between $T_c$ and $T^*$. To illustrate what happens in the system at these temperature points we included SAPNS  maps in Fig.\ref{fig:PPMS+SANS}(b-e).

In low $T$ region Bragg peaks evidence on the spiral magnetic structure (Fig.~\ref{fig:PPMS+SANS}(b)), which disappear at $T_{c}$ coinciding with temperature of maximum of $d\chi/dT$ in Fig.~\ref{fig:PPMS+SANS}(a). Just above $T_c$ the diffuse critical scattering of the chiral fluctuations is formed as a half moon, centred at the structure position (Fig.~\ref{fig:PPMS+SANS} (c)), similar to those observed in ref. \cite{GrigorievPRB05_v72}. This scattering  depends strongly on the initial polarization.  The asymmetric $P$-dependent scattering is a fingerprint of the single chirality of the spin helix that corresponds to the sign of the DMI. The radius of the ring of intensity observed above $T_c$ gives the value of the helix wave-vector $|\mathbf{ k}|$ and its width gives the inverse correlation length $\kappa$. The moon-like scattering image becomes blurred when temperature increases, $T > T_k$  (Fig.~\ref{fig:PPMS+SANS}(d)). Further increase of $T$ above $T^{*}$ makes the moon-like scattering to dissolve (Fig.~\ref{fig:PPMS+SANS}(e)). The quantitative analysis of the data is performed on the base of the mean-field theory of the critical neutron scattering for the cubic magnet with DMI \cite{GrigorievPRB05_v72, GrigorievPRB10_v81_2}. 

The neutron cross section for the paramagnetic phase ($T > T_\mathrm{c}$) is given in the following form:
\begin{eqnarray}
\frac{d\sigma}{d\Omega} = \frac{r^2T}{A} \;\frac{k^2 + \kappa^2 + Q^2 +2 \mathrm{\ sgn}\left(D\right)k\mathbf{Q} \cdot \mathbf{P}_0}{\left[\left(Q+k\right)^2+\kappa^2\right]\left[\left(Q - k\right)^2+\kappa^2\right]}.
\label{eq:CrossSection}
\end{eqnarray}
Here $\kappa$ is the inverse correlation length of the spin fluctuations, $\mathbf{Q}$ is the scattering vector, $A$ is the spin wave stiffness at low $T$, $k=2\pi/d$ is the helical wave vector ($d$ is the spiral period), $D$ is the Dzyaloshinskii constant, $\mathbf{P}_0$ is the incident polarization of neutrons, $T$ is the temperature. In this expression we have neglected the weak anisotropic exchange, which is essential in the very vicinity of $T_c$ only. Moreover, for doped compounds the local disorder smears the cubic anisotropy. The full description can be found in \cite{GrigorievPRB05_v72, GrigorievPRB10_v81_2}.

\begin{figure}[hbtp]
\begin{center}\includegraphics[width=8.5cm]{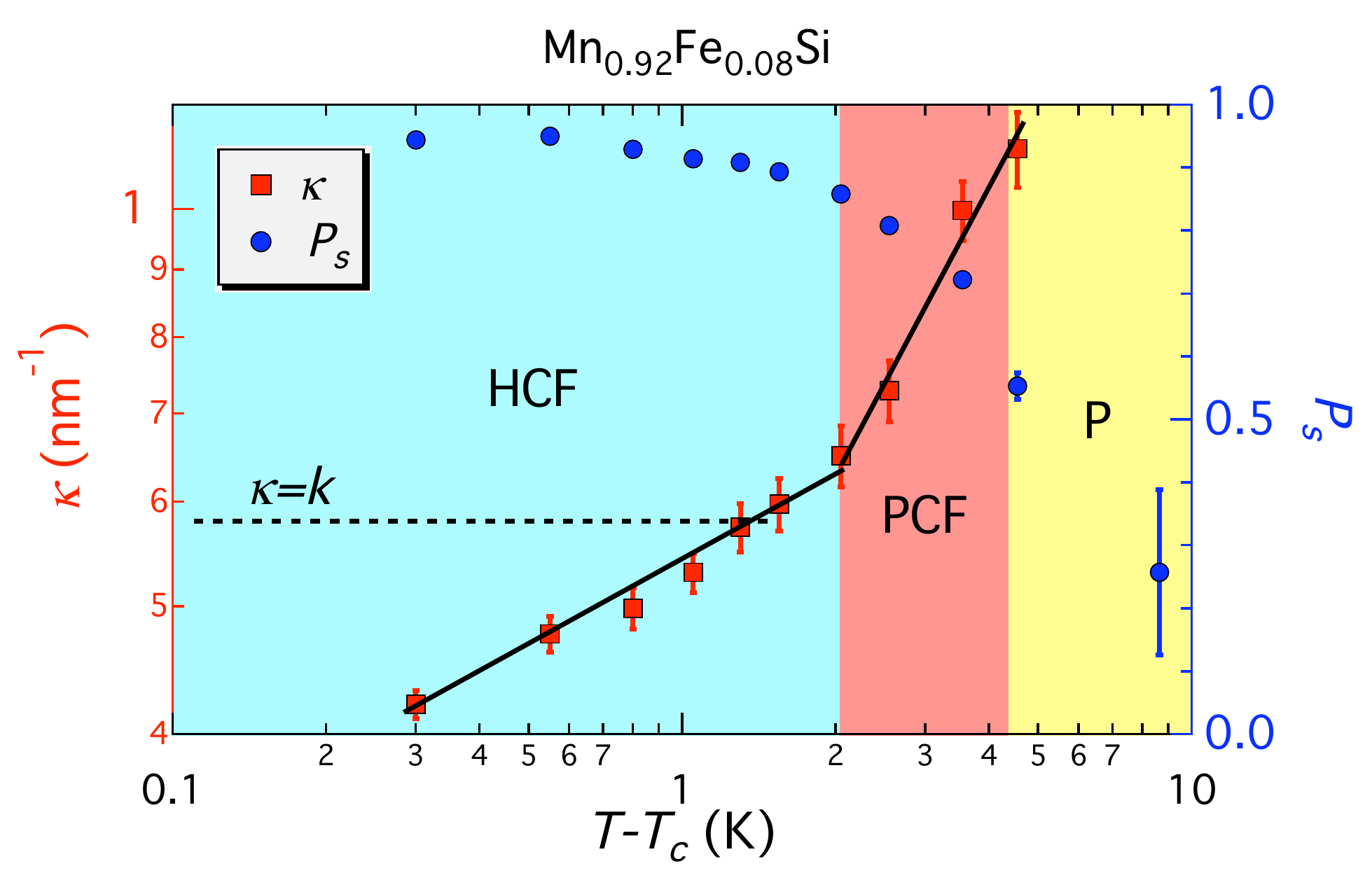}
\caption{The inverse correlation length $\kappa$ (squares) and the polarization $P_s$ (circles) as a function of the temperature $T-T_c$ for Mn$_{0.92}$Fe$_{0.08}$Si. Straight lines are guides to the eyes.}
\label{fig:mnfekappa}
\end{center}
\end{figure}
The cross section (Eq. \ref{eq:CrossSection}) has several features. (i) Due to the second term in the product in the denominator the scattering intensity should form a sphere of radius $Q = k$ with a width of $\kappa$. The scalar not vector variables in $(Q-k)^{2}$ is a result of isotropic DMI in cubic ferromagnets. (ii) The intensity has the Lorentzian shape typical for critical fluctuations. (iii) The scattering intensity is strongly oriented along the incident neutron polarization since the numerator of Eq.(\ref{eq:CrossSection}) is maximal at ${\bf P}_0 \uparrow\uparrow {\bf Q}$ and it is minimal at ${\bf P}_0 \uparrow\downarrow {\bf Q}$. The DMI is responsible for this kind of scattering. 

One can estimate the chirality of the critical fluctuations via measurement of the polarization of the scattering:
\begin{equation} 
P_s(Q) =\frac{\sigma({\bf P}_0)-\sigma(-{\bf P}_0)}{\sigma({\bf P}_0)+\sigma(-{\bf P}_0)}=-\frac {2 k Q P_0 \cos\phi}{Q^2+k^2+\kappa^2},
\label{eq:Pol}
\end{equation}
where $\phi$ is the angle between  ${\bf P}_0$ and ${\bf Q}$ and we use Eq.(\ref{eq:CrossSection}) to estimate its value via parameters $Q, k$ and $\kappa$.

The whole set of the experimental data obtained by polarized SANS was fitted to Eq.(\ref{eq:CrossSection}). The inverse correlation length $\kappa$ and the position of the maximum $k$ were obtained as a result of the fit.
The temperature dependence of the inverse correlation length $\kappa$ for Mn$_{0.92}$Fe$_{0.08}$Si is shown in Fig.~\ref{fig:mnfekappa}. The meaningful values for $\kappa$ had been possible to extract from the data in the temperature range $T_c < T < T^*$ (corresponding to the middle area in Fig. \ref{fig:PPMS+SANS}(a)), where $\kappa$ changes from $2k$ at $T = T^*$ to $(2/3) k$ at $T_c$. The temperature dependence of $\kappa$ in the log-log scale (Fig.~\ref{fig:mnfekappa}) exhibits a crossover at $T =T_k$, when $\kappa \simeq k$.

\begin{figure}[hbtp]
\begin{center}\includegraphics[width=8.5cm,keepaspectratio]{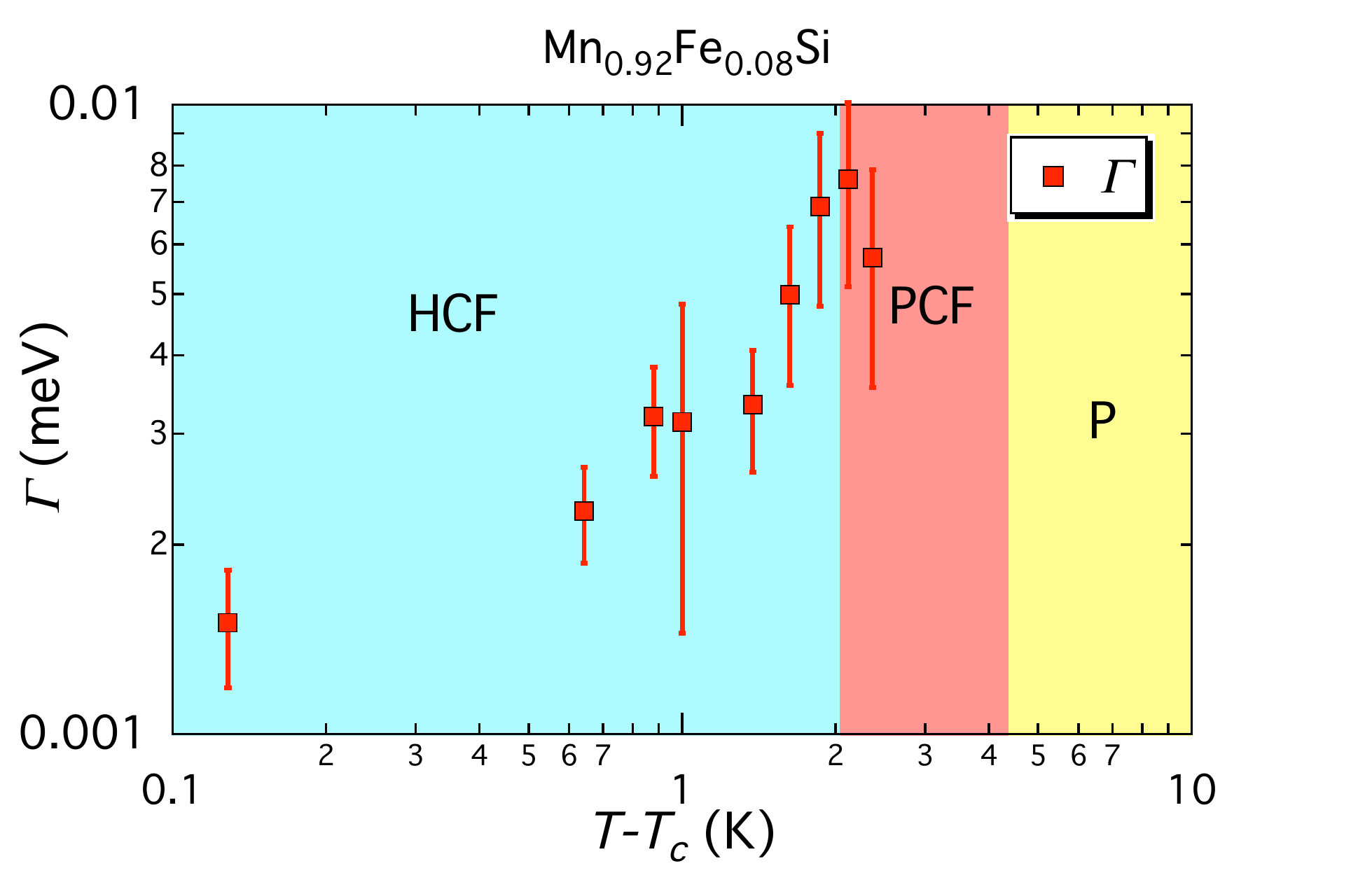}
\caption{The relaxation rate of the fluctuations, $\Gamma$, as a function of the temperature $T-T_c$ for Mn$_{0.92}$Fe$_{0.08}$Si.}
\label{fig:mnfeNSE}
\end{center}
\end{figure}

As it is well seen in  Fig.~\ref{fig:mnfekappa} and was also noted in 
\cite{GrigorievPRB05_v72,PappasPRL09_v102,GrigorievPRB10_v81_2} the polarization $P_s(Q =k)$ is close to 1 in the range  $T_c \leq T \leq T_k$, showing highly chiral fluctuating state within this range.  The polarization $P_s$ decreases smoothly upon increase of temperature above $T_k$. We refer to this change in the temperature behavior of $P_s$ as to the crossover from highly chiral fluctuating phase with $\kappa < k$ to the partially chiral phase with $\kappa>k$. Thus, this crossover in $T$-dependence of $P_s$ is well correlated to the crossover behavior of $\kappa (T)$.  

To study the critical dynamics in these compounds we have used the paramagnetic Neutron Spin Echo (NSE) technique. Experiments were performed on the wide angle NSE spectrometer SPAN at HZB in the same $Q$ range as was done for SANS. The temperature stabilization accuracy was $\Delta T=0.01$~K. All NSE spectra were normalized by the resolution function of the spectrometer, determined at $T=2$~K well below $T_{c}$, where there is no inelastic contribution to the spin-echo signal visible by this technique. The intermediate scattering function $S(Q, t)$ was measured at the magnetic Bragg point $\mathbf{Q} = {\bf k}\sim <111>$ within the chiral fluctuating phase $T_c \leq T \leq T_k$. The temperature dependence of the relaxation rate, $\Gamma$, is shown in Fig.~\ref{fig:mnfeNSE}.
As it can be seen from the picture, the relaxation rate of the critical fluctuations depends weakly on the temperature in the range of the highly chiral fluctuating phase for the given sample.


\label{Ph_Diag}

\begin{figure}[hbtp]
\begin{center}\includegraphics[width=8.5cm,keepaspectratio]{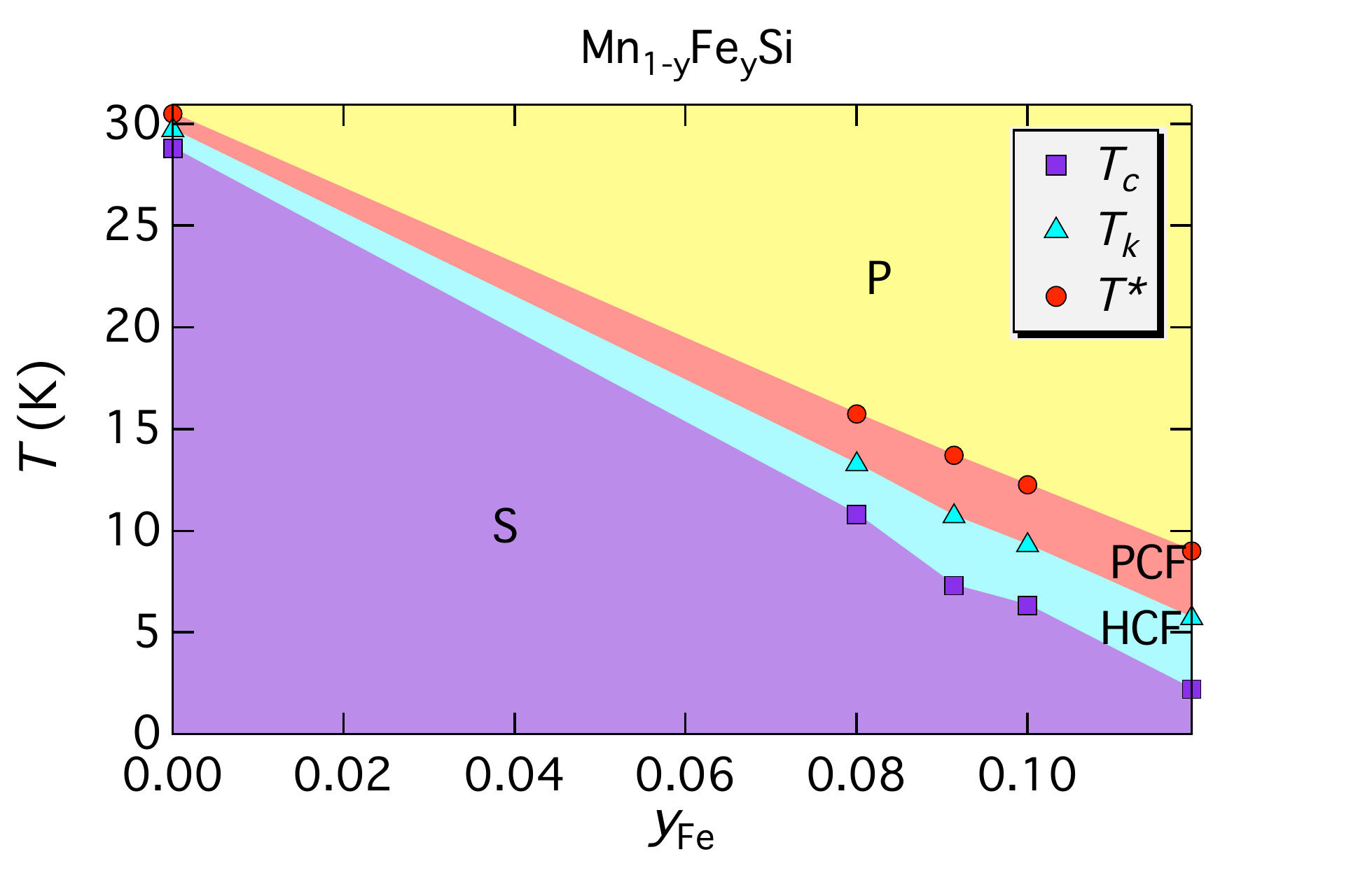}
\caption{Concentration dependencies of the critical temperature $T_\mathrm{c}$ and the temperatures of crossovers to PCF state $T^*$ and to HCF state $T_k$.}
\label{fig:CritTemp}
\end{center}
\end{figure}

{\it $(T-y)$ phase diagram.} From the ac-susceptibility measurements we have established three specific temperature points ($T_c$, $T_k$ and $T^*$), which are plotted in Fig.~\ref{fig:CritTemp}.

 The compounds undergo the transition from the paramagnetic (P) phase (with Curie-Weiss dependence at $T > T^*$) to spiral (S) phase ($T<T_c$) through two intermediate regions of Partially-Chiral Fluctuating (PCF) phase ($T_k<T<T^*$) and Highly-Chiral Fluctuating (HCF) phase ($T_c \leq T \leq T_k$) (Fig.~\ref{fig:CritTemp}). We have to stress that the transition is only one at $T=T_{c}$, and the other temperature points correspond to crossovers.
 
\begin{figure}[hbtp]
\begin{center}\includegraphics[width=8.5cm,keepaspectratio]{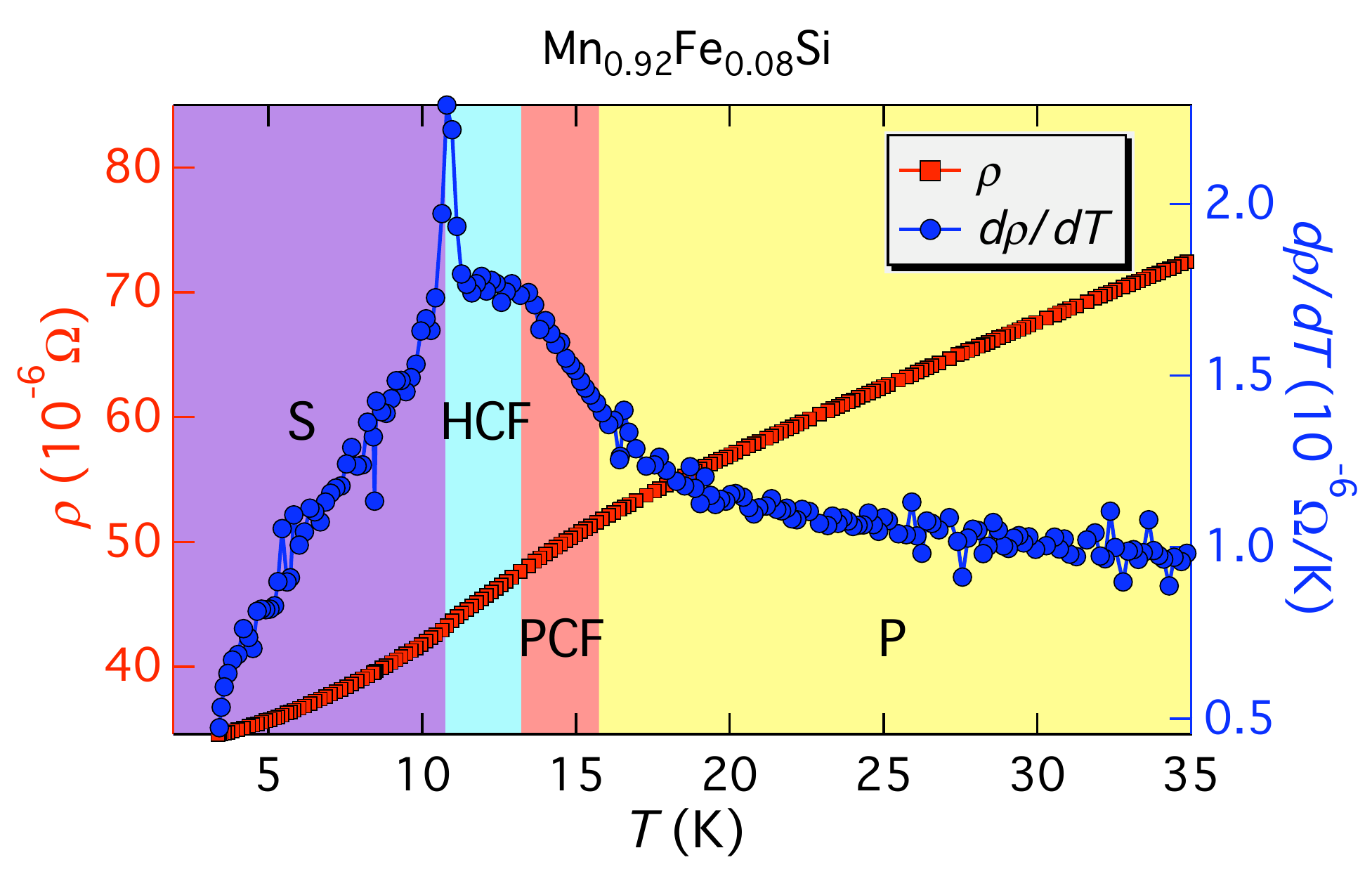}
\caption{The temperature dependence of resistivity, $\rho$, and its first derivative $d\rho/dT$, for Mn$_{0.92}$Fe$_{0.08}$Si.}
\label{fig:mnfeResist}
\end{center}
\end{figure}

Importance of the proposed delimitation of the ($T-y$) phase diagram is demonstrated in the temperature dependencies of the other, non magnetic physical properties of these compounds. Fig.~\ref{fig:mnfeResist} shows the temperature dependence of the resistivity of the compound with $y=0.08$. The first derivative $d\rho/dT$ enlightens the changes undergoing with the resistivity upon the temperature change. The critical temperature $T_c$ is marked by a sharp peak of $d\rho/dT$. A plato in $T$-dependence appears between $T_c$ and $T_k$, i.e. in the HCF state. The function decreases linearly in the PCF state in the range between $T_k$ and $T^*$. Thus, the shape of the $T$-dependence is very well correlated with four regions in the magnetic properties of the compounds. Interesting to note that this $T$-dependence duplicates temperature behaviour of the specific heat and thermal expansion coefficient \cite{StishovPRB07_v76}.

Accounting for the importance of the HCF state between $T_c$ and $T_k$ for the physical properties of these compounds we have collected in Table~\ref{tab:Params} the characteristic values of the inverse correlation length \begin{table}[htbt]
  \begin{center}
    \caption{\label{tab:Params} Parameters of Mn$_{1-y}$Fe$_{y}$Si compounds for the HCF state.}
      \begin{tabular}{ccccc}
        \hline
         \strut$y$ & $ T_c$(K) & $k$(nm$^{-1}$)    & $\kappa$(nm$^{-1}$) & $\Gamma$($\mu$eV)\\       \hline
           0       &  $29.0(1)$  &$0.40$    & $0.15 \leq \kappa \leq 0.3$ & 1 -- 2\\
           0.08    &  $10.8(4)$  &$0.58$    & $0.4 \leq \kappa \leq 0.6$  & 2 -- 5\\
           0.10    &  $6.5(5) $  & $0.64$   & $0.5 \leq \kappa \leq 0.7$  & 2 -- 4\\
      \end{tabular}
  \end{center}
\end{table}
$\kappa$ and energy $\Gamma$ of the critical fluctuations for different concentrations $y$. As seen in Table I, the critical temperature $T_c$ decreases with concentration $y$ while the spiral wavevector $k$ increases. These two quantities are interconnected via the two main energy scales: the DMI and the ferromagnetic exchange interaction (see for details \cite{Grigoriev09PRB_v79}).  The inverse correlation length, $\kappa$, within the HCF phase is clearly restricted by $\kappa_{min}(T_c) \leq \kappa \leq \kappa_{max}(T_k)\simeq k$. The relaxation rate $\Gamma$ changes weakly within the HCF state for an individual compound. For all compounds under study it can be satisfactory described by following empirical relation $\Gamma \sim T_c (k a)^x $ with $x \approx 4$. However, the dynamical properties of these compound in the critical range has not been studied theoretically yet.

{\it Conclusion.} The scenario and nature of the phase transition in the Mn$_{1-y}$Fe$_{y}$Si compounds are well described by comparison of the inverse correlation length $\kappa$ and the spiral wavevector $k$. In the high temperature range ($T>T^*$) the critical fluctuations are limited by $\kappa > 2k$, or $\xi < \pi/k = d/2$, where $d$ is the spiral period. For these fluctuations non-collinearity is not essential and they are close to the fluctuations of the conventional ferromagnet. In turn, the ac-susceptibility demonstrates Curie-Weiss behaviour  at $T>T^*$. The transition through the inflection point of susceptibility at $T^*$ lead to the PCF state with $k < \kappa < 2k$, or with the correlation length in real space $d/2 < \xi < d$. The full $2\pi$ twist of the helix is not completed yet inside these fluctuations. This results in rather low degree of chirality of such fluctuations. Further down on temperature ($T_c <T<T_k$), when the full $2\pi$ twist of the helix is well established within the fluctuations ($\kappa < k$), the highly chiral fluctuating phase appears. The temperature dependencies of both $\kappa$ and $P_s$ demonstrate the crossover at $\kappa \sim k$.  The crossover is compatible with the theory \cite{GrigorievPRB10_v81_2}. The global transition is completed at $T_c$, where the solid spiral structure arises. It is clear that if one neglect the weak crystal anisotropy, then the last transition is of the first order. This scenario with two crossovers at $T^*$, $T_k$ and transformation at $T_c$ is clearly reflected in the nonmagnetic properties of these compounds, such as resistivity, etc.   

The PNPI-team acknowledges GKSS for the hospitality. The work is partly supported by the RFBR projects No 09-02-01023-à, 10-02-01205-a and Goskontrakt No 02.740.11.0874.



\begin{thebibliography}{13}
\expandafter\ifx\csname natexlab\endcsname\relax\def\natexlab#1{#1}\fi
\expandafter\ifx\csname bibnamefont\endcsname\relax
  \def\bibnamefont#1{#1}\fi
\expandafter\ifx\csname bibfnamefont\endcsname\relax
  \def\bibfnamefont#1{#1}\fi
\expandafter\ifx\csname citenamefont\endcsname\relax
  \def\citenamefont#1{#1}\fi
\expandafter\ifx\csname url\endcsname\relax
  \def\url#1{\texttt{#1}}\fi
\expandafter\ifx\csname urlprefix\endcsname\relax\def\urlprefix{URL }\fi
\providecommand{\bibinfo}[2]{#2}
\providecommand{\eprint}[2][]{\url{#2}}

\bibitem[{\citenamefont{Nishihara et~al.}(1984)\citenamefont{Nishihara, Waki,
  and Ogawa}}]{NishiharaPRB84_v30}
\bibinfo{author}{\bibfnamefont{Y.}~\bibnamefont{Nishihara}},
  \bibinfo{author}{\bibfnamefont{S.}~\bibnamefont{Waki}}, \bibnamefont{and}
  \bibinfo{author}{\bibfnamefont{S.}~\bibnamefont{Ogawa}},
  \bibinfo{journal}{Phys. Rev. B} \textbf{\bibinfo{volume}{30}},
  \bibinfo{pages}{32} (\bibinfo{year}{1984}).

\bibitem[{\citenamefont{Grigoriev et~al.}(2009)\citenamefont{Grigoriev,
  Dyadkin, Moskvin, Lamago, Wolf, Eckerlebe, and Maleyev}}]{Grigoriev09PRB_v79}
\bibinfo{author}{\bibfnamefont{S.~V.} \bibnamefont{Grigoriev}},
  \bibinfo{author}{\bibfnamefont{V.~A.} \bibnamefont{Dyadkin}},
  \bibinfo{author}{\bibfnamefont{E.~V.} \bibnamefont{Moskvin}},
  \bibinfo{author}{\bibfnamefont{D.}~\bibnamefont{Lamago}},
  \bibinfo{author}{\bibfnamefont{T.}~\bibnamefont{Wolf}},
  \bibinfo{author}{\bibfnamefont{H.}~\bibnamefont{Eckerlebe}},
  \bibnamefont{and} \bibinfo{author}{\bibfnamefont{S.~V.}
  \bibnamefont{Maleyev}}, \bibinfo{journal}{Phys. Rev. B}
  \textbf{\bibinfo{volume}{79}}, \bibinfo{pages}{144417}
  (\bibinfo{year}{2009}).

\bibitem[{\citenamefont{Bak and Jensen}(1980)}]{BakJPCM80_v13}
\bibinfo{author}{\bibfnamefont{P.}~\bibnamefont{Bak}} \bibnamefont{and}
  \bibinfo{author}{\bibfnamefont{M.~H.} \bibnamefont{Jensen}},
  \bibinfo{journal}{J. Phys. C: Solid State Phys.}
  \textbf{\bibinfo{volume}{13}}, \bibinfo{pages}{L881} (\bibinfo{year}{1980}).

\bibitem[{\citenamefont{Dzyaloshinskii}(1964)}]{Dzyaloshinskii64ZETF}
\bibinfo{author}{\bibfnamefont{I.~E.} \bibnamefont{Dzyaloshinskii}},
  \bibinfo{journal}{Zh. Eksp. Teor. Fiz.} \textbf{\bibinfo{volume}{46}},
  \bibinfo{pages}{1420} (\bibinfo{year}{1964}).

\bibitem[{\citenamefont{Nakanishi et~al.}(1980)\citenamefont{Nakanishi, Yanase,
  Hasegawa, and Kataoka}}]{NakanishiSSC80_v35}
\bibinfo{author}{\bibfnamefont{O.}~\bibnamefont{Nakanishi}},
  \bibinfo{author}{\bibfnamefont{A.}~\bibnamefont{Yanase}},
  \bibinfo{author}{\bibfnamefont{A.}~\bibnamefont{Hasegawa}}, \bibnamefont{and}
  \bibinfo{author}{\bibfnamefont{M.}~\bibnamefont{Kataoka}},
  \bibinfo{journal}{Solid State Commun.} \textbf{\bibinfo{volume}{35}},
  \bibinfo{pages}{995} (\bibinfo{year}{1980}).

\bibitem[{\citenamefont{Roessli et~al.}(2002)\citenamefont{Roessli, B\"{o}ni,
  Fischer, and Endoh}}]{RoessliPRL02_v88}
\bibinfo{author}{\bibfnamefont{B.}~\bibnamefont{Roessli}},
  \bibinfo{author}{\bibfnamefont{P.}~\bibnamefont{B\"{o}ni}},
  \bibinfo{author}{\bibfnamefont{W.~E.} \bibnamefont{Fischer}},
  \bibnamefont{and} \bibinfo{author}{\bibfnamefont{Y.}~\bibnamefont{Endoh}},
  \bibinfo{journal}{Phys. Rev. Lett.} \textbf{\bibinfo{volume}{88}},
  \bibinfo{pages}{237204} (\bibinfo{year}{2002}).

\bibitem[{\citenamefont{Grigoriev et~al.}(2005)\citenamefont{Grigoriev,
  Maleyev, Okorokov, Chetverikov, Georgii, Boni, Lamago, Eckerlebe, and
  Pranzas}}]{GrigorievPRB05_v72}
\bibinfo{author}{\bibfnamefont{S.~V.} \bibnamefont{Grigoriev}},
  \bibinfo{author}{\bibfnamefont{S.~V.} \bibnamefont{Maleyev}},
  \bibinfo{author}{\bibfnamefont{A.~I.} \bibnamefont{Okorokov}},
  \bibinfo{author}{\bibfnamefont{Y.~O.} \bibnamefont{Chetverikov}},
  \bibinfo{author}{\bibfnamefont{R.}~\bibnamefont{Georgii}},
  \bibinfo{author}{\bibfnamefont{P.}~\bibnamefont{Boni}},
  \bibinfo{author}{\bibfnamefont{D.}~\bibnamefont{Lamago}},
  \bibinfo{author}{\bibfnamefont{H.}~\bibnamefont{Eckerlebe}},
  \bibnamefont{and} \bibinfo{author}{\bibfnamefont{K.}~\bibnamefont{Pranzas}},
  \bibinfo{journal}{Phys. Rev. B} \textbf{\bibinfo{volume}{72}},
  \bibinfo{pages}{134420} (\bibinfo{year}{2005}).

\bibitem[{\citenamefont{R\"ossler et~al.}(2006)\citenamefont{R\"ossler,
  Bogdanov, and Pfleiderer}}]{RoesslerNature06_v442}
\bibinfo{author}{\bibfnamefont{U.~K.} \bibnamefont{R\"ossler}},
  \bibinfo{author}{\bibfnamefont{A.~N.} \bibnamefont{Bogdanov}},
  \bibnamefont{and}
  \bibinfo{author}{\bibfnamefont{C.}~\bibnamefont{Pfleiderer}},
  \bibinfo{journal}{Nature} \textbf{\bibinfo{volume}{442}},
  \bibinfo{pages}{797} (\bibinfo{year}{2006}).

\bibitem[{\citenamefont{Stishov et~al.}(2007)\citenamefont{Stishov, Petrova,
  Khasanov, Panova, Shikov, Lashley, Wu, and Lograsso}}]{StishovPRB07_v76}
\bibinfo{author}{\bibfnamefont{S.~M.} \bibnamefont{Stishov}},
  \bibinfo{author}{\bibfnamefont{A.~E.} \bibnamefont{Petrova}},
  \bibinfo{author}{\bibfnamefont{S.}~\bibnamefont{Khasanov}},
  \bibinfo{author}{\bibfnamefont{G.~K.} \bibnamefont{Panova}},
  \bibinfo{author}{\bibfnamefont{A.~A.} \bibnamefont{Shikov}},
  \bibinfo{author}{\bibfnamefont{J.~C.} \bibnamefont{Lashley}},
  \bibinfo{author}{\bibfnamefont{D.}~\bibnamefont{Wu}}, \bibnamefont{and}
  \bibinfo{author}{\bibfnamefont{T.~A.} \bibnamefont{Lograsso}},
  \bibinfo{journal}{Phys. Rev. B} \textbf{\bibinfo{volume}{76}},
  \bibinfo{pages}{052405} (\bibinfo{year}{2007}).

\bibitem[{\citenamefont{Pappas et~al.}(2009)\citenamefont{Pappas,
  Leli\`{e}vre-Berna, Falus, Bentley, Moskvin, Grigoriev, Fouquet, and
  Farago}}]{PappasPRL09_v102}
\bibinfo{author}{\bibfnamefont{C.}~\bibnamefont{Pappas}},
  \bibinfo{author}{\bibfnamefont{E.}~\bibnamefont{Leli\`{e}vre-Berna}},
  \bibinfo{author}{\bibfnamefont{P.}~\bibnamefont{Falus}},
  \bibinfo{author}{\bibfnamefont{P.~M.} \bibnamefont{Bentley}},
  \bibinfo{author}{\bibfnamefont{E.}~\bibnamefont{Moskvin}},
  \bibinfo{author}{\bibfnamefont{S.}~\bibnamefont{Grigoriev}},
  \bibinfo{author}{\bibfnamefont{P.}~\bibnamefont{Fouquet}}, \bibnamefont{and}
  \bibinfo{author}{\bibfnamefont{B.}~\bibnamefont{Farago}},
  \bibinfo{journal}{Phys. Rev. Lett.} \textbf{\bibinfo{volume}{102}},
  \bibinfo{pages}{197202} (\bibinfo{year}{2009}).

\bibitem[{\citenamefont{Grigoriev et~al.}(2010)\citenamefont{Grigoriev,
  Maleyev, Moskvin, Dyadkin, Fouquet, and Eckerlebe}}]{GrigorievPRB10_v81_2}
\bibinfo{author}{\bibfnamefont{S.~V.} \bibnamefont{Grigoriev}},
  \bibinfo{author}{\bibfnamefont{S.~V.} \bibnamefont{Maleyev}},
  \bibinfo{author}{\bibfnamefont{E.~V.} \bibnamefont{Moskvin}},
  \bibinfo{author}{\bibfnamefont{V.~A.} \bibnamefont{Dyadkin}},
  \bibinfo{author}{\bibfnamefont{P.}~\bibnamefont{Fouquet}}, \bibnamefont{and}
  \bibinfo{author}{\bibfnamefont{H.}~\bibnamefont{Eckerlebe}},
  \bibinfo{journal}{Phys. Rev. B} \textbf{\bibinfo{volume}{81}},
  \bibinfo{pages}{144413} (\bibinfo{year}{2010}).

\bibitem[{\citenamefont{Tewari et~al.}(2006)\citenamefont{Tewari, Belitz, and
  Kirkpatrick}}]{TewariPRL06_v96}
\bibinfo{author}{\bibfnamefont{S.}~\bibnamefont{Tewari}},
  \bibinfo{author}{\bibfnamefont{D.}~\bibnamefont{Belitz}}, \bibnamefont{and}
  \bibinfo{author}{\bibfnamefont{T.~R.} \bibnamefont{Kirkpatrick}},
  \bibinfo{journal}{Phys. Rev. Lett.} \textbf{\bibinfo{volume}{96}},
  \bibinfo{pages}{047207} (\bibinfo{year}{2006}).

\bibitem[{\citenamefont{Bauer et~al.}(2010)\citenamefont{Bauer, Neubauer,
  Franz, M\"unzer, Garst, and Pfleiderer}}]{BauerPRB10_v82}
\bibinfo{author}{\bibfnamefont{A.}~\bibnamefont{Bauer}},
  \bibinfo{author}{\bibfnamefont{A.}~\bibnamefont{Neubauer}},
  \bibinfo{author}{\bibfnamefont{C.}~\bibnamefont{Franz}},
  \bibinfo{author}{\bibfnamefont{W.}~\bibnamefont{M\"unzer}},
  \bibinfo{author}{\bibfnamefont{M.}~\bibnamefont{Garst}}, \bibnamefont{and}
  \bibinfo{author}{\bibfnamefont{C.}~\bibnamefont{Pfleiderer}},
  \bibinfo{journal}{Phys. Rev. B} \textbf{\bibinfo{volume}{82}},
  \bibinfo{pages}{064404} (\bibinfo{year}{2010}).

\end{thebibliography}

\end{document}